\documentclass[prd,amsfonts,amssymb,
amsmath,showkeys,showpacs]{revtex4}

\date{\today}
\usepackage{graphicx}

\begin{document}

\author{Leonard Parker}
\email{leonard@uwm.edu}
\affiliation{Physics Department, University of Wisconsin-Milwaukee,
P.O.Box 413, Milwaukee, Wisconsin USA 53201}

\title{Amplitude of Perturbations from Inflation}

\begin{abstract}
The observed power spectrum of the cosmic microwave 
background (CMB) is consistent with inflationary cosmology, 
which predicts a nearly scale-invariant power spectrum of 
quantum fluctuations of the inflaton field as 
they exit the Hubble horizon during inflation. Here we report
a very significant correction (of several orders of magnitude) to the 
predicted amplitude of the power spectrum. This correction does not 
alter the near scale-invariance of the spectrum, but is crucial for testing
predictions of the Hubble parameter during inflation against the 
observed amplitude of the CMB power spectrum.
This novel correction appears because, as we show, the subtractions
that renormalize the short-wavelength ultraviolet divergences of the
inflaton two-point function have a significant effect on the
amplitude of that two-point function at the longer wavelengths
characteristic of the Hubble horizon. Earlier conclusions in the literature
that certain theories (such as grand unified theories) implied
perturbations that were too large by several orders of magnitude
will have to be reconsidered in light of the present result.
\pacs{98.80.Cq, 04.62.+v, 98.80.Es, 98.70.Vc}
% particle and field theory models of early universe-inflation
% quantum field theory in curved spacetime
% observational cosmology
% background radiations
%
\keywords{inflation; spectrum of perturbations; CMB;
                        quantum field theory in curved spacetime}
\end{abstract}
\newpage

\maketitle

\section{Introduction}
\label{sec:introduction}

The primordial perturbations responsible for the anisotropies in 
the CMB \cite{PeirisWMAP1, SpergelWMAP1, SpergelWMAP3} 
are likely to have originated as quantum fluctuations of a scalar 
field $\phi$, the inflaton, that was the dominant constituent of the 
universe during a period of rapid exponential inflation
\cite{Guth81, Linde82, AlbrechtSteinhardt82, Sato81, MukhanovChibisov81,
Hawking82, GuthPi82, Starobinsky82, BardeenSteinhardtTurner83} 
that lasted long enough to allow 
all parts of our observed universe to be causally connected. The 
potential-energy density of the inflaton field 
provided the impetus for the inflationary expansion, while its quantum
fluctuations upon leaving the Hubble event horizon during inflation 
provided the initial conditions for the metric and density perturbations 
at the beginning of the radiation-dominated stage of the expansion 
(see, for example, the books \cite{LiddleLyth2000, Dodelson2003}). We consider 
the following typical slow-roll inflationary scenario, which is consistent 
with the three-year Wilkinson Microwave Anisotropy Probe (WMAP)
observations \cite{SpergelWMAP3}. As first shown by the 
author \cite{Parker66, Parker69}, 
the generation and amplification of quantum fluctuations of a scalar field,
such as the inflaton, is inevitable during an expansion of the universe 
like that of inflation.

The metric during exponential inflation is
\begin{equation}
ds^2= dt^2-\exp(2H t) \left( (dx)^2 +(dy)^2+(dz)^2\right)
\label{eq:metric}
\end{equation}
with $H$ given by
\begin{equation}
H  = \sqrt{\frac{8 \pi G}{3} V(\phi^{(0)}) } \ .
\label{eq:HofV}
\end{equation}
where G is Newton's constant and $V(\phi^{(0)})$ is the inflaton potential. The 
zeroth-order inflaton field $\phi^{(0)}$ is treated classically.  It and $H$ change 
very gradually in slow-roll inflation as $\phi^{(0)}$ rolls slowly down the potential 
toward a minimum.

The inflaton field to first order is  written as
\begin{equation}
\phi(\vec{x}, t)=\phi^{(0)}(t)+\delta\phi(\vec{x}, t)
\label{eq:phiToFirstOrder}
\end{equation}
The first-order perturbation obeys the minimally-coupled linear scalar field equation on the background metric:
\begin{equation}
{\partial_t}^2\delta\phi + 3 H \partial_t \,\delta\phi -
        \exp(-2Ht)\sum_{i=1}^{3} {\partial_i}^2 \delta\phi+m(\phi^{(0)})^2\delta\phi=0
\label{eq:FirstOrderScalarFieldEq}
\end{equation}
where $m(\phi^{(0)})^2 \equiv V''(\phi^{(0)})$.
The conditions of slow-roll inflation imply that $V''\,/G \ll V$, from which it follows that
$m^2$ is small with respect to $H^2$. For inflationary potentials that are usually
considered and are consistent with the WMAP data, $m^2$ is positive and nearly
constant during inflation. This is a convenient framework in which to illustrate our 
main point, namely, that the subtractions required for the renormalization of 
ultraviolet (UV) divergences in the two-point function reduce by several orders 
of magnitude the amplitude of the inflaton fluctuations at the times when
their wavelengths leave the Hubble horizon, as compared with the results that
are generally found in the literature. 

The method of adiabatic 
regularization that determines the subtraction terms has been known
for many years \cite{Parker66, ParkerFulling74, ParkerFullingHu74, 
Bunch80, AndersonParker87,
AndersonEakerHabibParisMottola2000, Finelli05, AndersonParisMottola2005,
ParkerVanzella2007}. Below, we find that the adiabatic subtractions 
that bring the UV divergences of the
two-point function under control have a very significant effect on the
amplitude of the two-point function at the frequencies corresponding to
exit of the inflaton perturbations from the inflationary Hubble horizon. 
A pedagogical explanation of the adiabatic subtraction method
as it applies to the two-point function and the energy-momentum tensor
is given in the book \cite{ParkerToms2007} (see also \cite{BirrellDavies82}). 

Although it is
clear that inflation is not an adiabatic (slow) expansion of the universe,
nevertheless, from the point of view of the short-wavelength, short-period 
fluctuations that 
give rise to the UV divergences of the two-point function, the Hubble
period, $H^{-1}$, of the inflationary expansion is long, so the inflationary expansion
appears adiabatic at very high UV frequencies. Thus, adiabatic regularization
and subtraction is applicable to the two-point function of the
quantized inflaton perturbation field during inflation.
We show that the amplitudes of these perturbations will be reduced by
several orders of magnitude compared to their amplitudes without the
adiabatic subtractions. This effect is crucial when
considering fundamental physical theories that predict the value of
$H$ during inflation. It will change previous results in the
literature that showed that certain fundamental theories
seemed to give perturbations from inflation that were
too large by orders of magnitude.

The coordinate system $(x, y, z, t)$  that appears in $ds^2$ above covers a 
part of deSitter spacetime. The well-known normalized positive-frequency
solution of the above differential equation for the field $\delta \phi$ that is 
determined to within a constant phase factor by the deSitter isometries 
of spatial rotations and translations, and by the deSitter extension of time 
translation, is
\begin{equation}
f_{\vec{k}}(\vec{x}, t)=(2 L^3 a(t)^3)^{-1/2} h_{k}(t)\exp(i \vec{k}\cdot \vec{x})
\label{eq:fk}
\end{equation}
with $a(t)\equiv \exp(Ht)$, and
\begin{equation}
h_{k}(t)=\sqrt{\frac{\pi}{2H}}\, H^{(1)}_{n}(v)
\label{eq:hk}
\end{equation}
where $H^{1}_{n}(v)$ is a Hankel function of the first kind, and
\begin{equation}
n=\sqrt{\frac{9}{4}-\frac{m^2}{H^2}}\ \ \ \ {\rm and}\ \ \ v=k H^{-1} \exp(-Ht).
\label{eq:nv}
\end{equation}
In the present analysis, we are treating $H$ and $m$ as constant to first
approximation during inflation.
Note that $v=2\pi H^{-1}\, / \lambda(k, t)$ is the ratio of the radius of the Hubble 
horizon $H^{-1}$ to the wavelength of the inflaton perturbation
$\lambda(k, t)=2\pi a(t)/k$. For a given value of $k$, the convention is to
define the time of exit of the corresponding mode from the Hubble horizon
as the time when $v = 1$.

We have temporarily imposed periodic boundary 
conditions on the modes $f_{\vec{k}}(\vec{x}, t)$ in a spatial cube of volume 
$L^3$,  where $L$ is taken to infinity in the continuum limit. The hermitian 
quantized field $\delta\phi$ is expressed as usual in the mode expansion
\begin{equation}
\delta\phi(\vec{x}, t)=\sum_{\vec{k}}\, (A_{\vec{k}} f_{\vec{k}}(\vec{x}, t)+{\rm h.c.})
\label{eq:deltaphi}
\end{equation}
where \mbox{h.c.} denotes the hermitian conjugate of the previous term,
and $A_{\vec{k}}$ is an
annihilation operator for a particle of the quantized inflaton perturbation field in 
mode $f_{\vec{k}}(\vec{x}, t)$. The deSitter or Bunch-Davies  
vacuum state $\left|0\right>$ is defined by
 $A_{\vec{k}}\left|0\right>=0$ for all $\vec{k}$. We are working in the Heisenberg
 picture, and expectation values below are taken with respect to this state vector.
 This vacuum state is a natural one to take for the inflaton perturbation field
 in the exponentially expanding inflationary universe.

\section{Adiabatic  subtractions necessary for renormalization}
\label{sec:AdiabaticSubtractions}

The formal two-point function $\left<0\right| \delta\phi(\vec{x}, t)^2   \left|0\right>$
obtained from Eq.\ (\ref{eq:deltaphi}) is
\begin{equation}
\left<0\right | \left | \delta\phi(\vec{x}, t) \right |^2  \left |0\right>_{\rm formal}=
  \sum_{\vec{k}}\, \left |  f_{\vec{k}}(\vec{x}, t) \right |^2 =
  (4\pi a(t)^3)^{-1} \int_{0}^{\infty} \left | h_{k}(t) \right |^2 k^2 dk 
\label{eq:formal2point}
\end{equation}
where we have taken the continuum limit. From the large-k asymptotic form of
the Hankel function, one finds that this integral over k diverges quadratically at
the upper limit of integration. 

Well established methods have been developed to 
deal with such UV divergences that arise in quantum field theory in curved 
spacetime, as normal ordering of operators is not sufficient in an evolving 
universe. The method that best suits our present objective of finding the 
{\em spectrum} of perturbations is the method of adiabatic regularization first 
developed by Parker and Fulling 
\cite{Parker66, ParkerFulling74, ParkerFullingHu74,% 
Bunch80, AndersonParker87}.
The method involves a mode by mode subtraction of terms that are related to the 
behavior of the mode functions in a slowly (i.e., adiabatically) expanding 
universe. Adiabatic regularization is related to the fact that the particle number 
is an adiabatic invariant \cite{Parker66, Parker69} in a slowly 
expanding universe.  Even the rate of the inflationary expansion is slow 
relative to the high frequencies that give rise to the UV divergence. The
adiabatic subtractions are obtained from an adiabatic series for the
mode functions. They serve to identify and eliminate the UV divergences
that are present in quantities such as the two-point function and
expectation values of the energy-momentum tensor. The method is
very simple to apply in the Friedmann, Lemaitre, Robertson,
Walker (FLRW) background metrics, such as that of Eq.\ (\ref{eq:metric}). %1

The adiabatically regularized expression for the two-point function is
\begin{equation}
\left<0\right | \left | \delta\phi(\vec{x}, t) \right |^2  \left |0\right> =
    (4\pi a(t)^3)^{-1} \int_{0}^{\infty} \left( \left | h_{k}(t) \right |^2 -
       \omega_k(t)^{-1} - (W_k(t)^{-1})^{(2)} \right) k^2 dk 
\label{eq:renorm2point}
\end{equation}
where the first adiabatic subtraction term removes the quadratic UV divergence
and the second subtraction removes the logarithmic UV divergence that is
present in the formal expression of Eq.\ (\ref{eq:formal2point}).  The full
expression in the integrand is well-defined and gives a finite result, even
though if one were to break apart the separate integrals, they would each
suffer from a UV divergence. Here $\omega_k(t)=\sqrt{\frac{k^2}{a(t)^2}+m^2}$,
and for the present exponentially expanding universe and field equation, the
second subtraction term is 
\begin{equation}
(W_k(t)^{-1})^{(2)} = -\frac{5 H^2 m^4}{8\omega_k(t)^7}+
       \frac{3 H^2 m^2}{4\omega_k(t)^5} + \frac{H^2}{\omega_k(t)^3}.
\label{eq:W2subtraction}
\end{equation}
The adiabatic subtractions must be taken at all frequencies, not just large ones.
Otherwise fundamental properties would be violated; for example, the
covariant four-divergence of the renormalized energy-momentum tensor
would not be zero.

\section{The spectrum of inflaton perturbations}
\label{sec:SpectrumInflaton}

The two-point function is related to the spectrum of inflaton perturbations 
($\Delta_{\phi}^2(k,t)$, also denoted as ${\cal P}_\phi (k,t)$) by %2
\begin{equation}
\left<0\right | \left | \delta\phi(\vec{x}, t) \right |^2  \left |0\right> =
\int_{0}^{\infty} \Delta_{\phi}^2(k,t) k^{-1} dk
\label{eq:SpectrumAnd2point}
\end{equation}
Comparing with Eq.\ (\ref{eq:renorm2point}), one finds 
\begin{equation}
\Delta_{\phi}^2(k,t) =
    (4\pi a(t)^3)^{-1} k^3 \left( \left | h_{k}(t) \right |^2 -
       \omega_k(t)^{-1} - (W_k(t)^{-1})^{(2)} \right) 
\label{eq:SpectrumAndSubtractions}
\end{equation}
This is equivalent to the definition of the spectrum as \cite{PeirisWMAP1}:
\begin{equation}
\Delta_{\phi}^2(k,t)= k^3/(2\pi^2)
        \left<0\right | \left | \delta\phi_{\vec{k}}( t) \right |^2  \left |0\right> 
\label{eq:SpectrumAndMode2point}
\end{equation}
where we define the renormalized momentum-space 
expectation value as
\begin{equation} 
\left<0\right | \left | \delta\phi_{\vec{k}}( t) \right |^2  \left |0\right> =
       (2\pi^2)(4\pi a(t)^3)^{-1} \left( \left | h_{k}(t) \right |^2 -
       \omega_k(t)^{-1} - (W_k(t)^{-1})^{(2)} \right) .
\label{eq:RenormMode2point}
\end{equation}
As we have seen, consistency between the physically relevant renormalized
two-point functions in position space and momentum space demands that
the physically relevant perturbations in momentum space are given by 
Eq.\ (\ref{eq:SpectrumAndSubtractions}).

Having obtained the momentum-space two-point function of
Eq.\ (\ref{eq:RenormMode2point}) 
directly from the renormalized position-space two-point function, one
may ask if there is an alternative argument for the physical relevance
of the subtractions that deals with the individual momentum-space modes
in the expansion of the quantized field? Such an argument can be
found in the author's Ph.D. thesis \cite[pp. 140--163 ]{Parker66}, 
in which quantum measurement theory was
used to redefine the creation and annihilation operators that are measurable
for each mode in the expansion of the field analogous to Eq.\ (\ref{eq:deltaphi}).
The modes in the field expansion, when expressed in terms of these physically 
relevant creation and annihilation operators, would lead to the result for
the momentum-space two-point function given in 
Eq.\ (\ref{eq:RenormMode2point}), with the vacuum state being the one
annihilated by the physical annihilation operators. In the present case, the 
connection with quantum measurement theory can be thought of in
terms of the classical curvature perturbations that give the initial conditions
for the radiation-dominated stage of our universe. The classical curvature
perturbations that arise from the quantum inflaton perturbations 
can be regarded as making a measurement on them, with the readout
of this measuring instrument being the observed power spectrum of
the CMB.

Simply defining the spectrum $\Delta_{\phi}^2(k,t)$ by using the Fourier
components in the field expansion of Eq.\ (\ref{eq:deltaphi}) would give
for $\left<0\right | \left | \delta\phi_{\vec{k}}( t) \right |^2  \left |0\right>$ the
result on the right-hand-side of Eq.\ (\ref{eq:RenormMode2point}), but
{\em without} any of the adiabatic subtractions. 
Then the position space two-point function 
$\left<0\right | \left | \delta\phi(\vec{x}, t) \right |^2  \left |0\right>$
obtained from  Eq.\ (\ref{eq:SpectrumAnd2point}) would be infinite. 
One may try to tame this infinity by the {\it ad hoc} prescription of 
demanding that $\delta\phi_{\vec{k}}( t)$ approach $0$ 
rapidly at sufficiently large $k$.  In this procedure, one obtains
for the modes of interest:
\begin{equation}
\Delta_{\phi}^2(k,t)_{\rm Hankel} \equiv
    (4\pi a(t)^3)^{-1} k^3 \left | h_{k}(t) \right |^2  
\label{eq:AdHocSpectrum1}
\end{equation}
However, this {\it ad hoc} 
procedure does not correctly take into account the fact that the physically
relevant spectrum of Eq.\ (\ref{eq:SpectrumAndSubtractions}) is
significantly different from the {\it ad hoc} spectrum obtained by this 
cut-off procedure.

Another often used procedure\cite{LiddleLyth2000} is to combine the 
ad hoc cut-off idea with a method of putting boundary conditions on the
mode functions such that the end result is to find the same result as
Eq.\ (\ref{eq:AdHocSpectrum1}), but with 
$\left | h_{k}(t) \right |^2= \omega_k(t)^{-1}$. This gives, for the
modes of interest, the spectrum
\begin{equation}
\Delta_{\phi}^2(k,t)_{\omega} \equiv
    (4\pi a(t)^3)^{-1} k^3 \omega_k(t)^{-1}.
\label{eq:AdHocSpectrum2}
\end{equation}
Without the cut-off, this would also give an infinite position-space
two-point function. 

At horizon exit ($v = 1$) it turns out that for small
values of $m^2$, the last two spectra are related by
$\Delta_{\phi}^2(k,t)_{\rm Hankel} \approx
2 \Delta_{\phi}^2(k,t)_{\omega}$. It is our contention that neither of
these last two spectra are correct because they do not take account
of the subtraction terms  that are necessary for renormalization
in curved spacetime.

Next we compare the actual physically relevant spectrum with the two
{\it ad hoc} spectra. We will find that the near scale-independence 
of the spectral components at the times of exit from the Hubble horizon
is not altered, but the magnitudes are quite different. Thus, the 
{\em agreement} with CMB observations 
is not altered. But for comparison of the observations with
theories that predict the magnitude of the Hubble parameter
$H$ during inflation, the difference in amplitude between the
actual spectrum and the two {\it ad hoc} spectra is quite important.

\section{Comparison of spectral amplitudes}
\label{sec:Comparison}

As noted, exit from the Hubble horizon is defined in the literature as occurring 
for each inflaton perturbation mode at the time when $v = 1$, 
where $v$ is defined in Eq. (\ref{eq:nv}). Expressing each of the spectra
given by Eqs.\ (\ref{eq:SpectrumAndSubtractions}), 
(\ref{eq:AdHocSpectrum1}), and (\ref{eq:AdHocSpectrum2})
in terms of $v$, we find that the spectrum we are proposing is
\begin{equation}
\Delta_{\phi}^2(k,t) = \frac{H^2 v^3}{32\pi^2} 
       \left (  4\pi \left | H^{(1)}_{n}(v)  \right |^2 -  
                 \frac{8m_H^6 + 3m_H^4 (3 + 8v^2) + 
                          2m_H^2 v^2 (11+12v^2) + 8(v^4+v^6)}{(m_H^2 + v^2)^{7/2}}
        \right )
\label{eq:DeltaSquared}
\end{equation}
where $n=\sqrt{(9/4) - m_H^2}$ and $m_H = m/H$. The other two spectra 
are, respectively,
\begin{equation}
\Delta_{\phi}^2(k,t)_{\rm Hankel} = \frac{H^2 v^3}{8\pi} 
        \left | H^{(1)}_{n}(v)  \right |^2 
\label{eq:DeltaSquared-h}
\end{equation}
and
\begin{equation}
\Delta_{\phi}^2(k,t)_{\omega} = \frac{H^2 v^3}{4\pi^2\sqrt{v^2 + m_H^2}} 
\label{eq:DeltaSquared-omega}
\end{equation}
When $v=1$ and $m \ll H$, one has for the spectrum of 
Eq.\ (\ref{eq:DeltaSquared-omega}),
$\Delta_{\phi}^2(k,t)_{\omega} = \frac{H^2}{4\pi^2}$,
which is the result used in the literature 
\cite{PeirisWMAP1, SpergelWMAP1, SpergelWMAP3,% 
LiddleLyth2000, Dodelson2003}.
We are proposing that instead one should use the result 
obtained from Eq.\ (\ref{eq:DeltaSquared}).

\begin{figure}
  \includegraphics{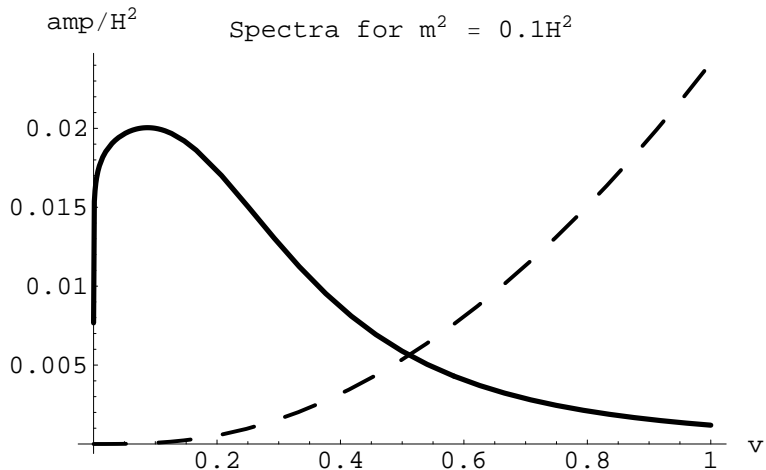}
  \caption{Plot of proposed spectrum (solid curve) of 
  Eq.\ (\ref{eq:DeltaSquared})
  and of the spectrum (dashed line) of 
  Eq.\ (\ref{eq:DeltaSquared-omega}), 
  for
  $m^2=0.1H^2$. The amplitude is in units of $H^2$. The quantity
  $v=k H^{-1} \exp(-Ht)$ is $1$ at horizon exit.}
  \label{fig1}
\end{figure}

All three of these spectra depend on the mode $k$ only through
the variable $v$. Therefore, the spectra obtained from the modes 
as they exit the Hubble horizon will  be scale-invariant in each case 
and will give rise to scale-invariant initial curvature perturbations 
at the beginning of the radiation-dominated
stage of the expansion. (Because $H$ depends on the unperturbed
classical inflaton field $\phi^{(0)}$, which is slowly rolling down the
inflaton potential, there will be some deviation from the scale-invariant
spectrum, as is observed  and has been
thoroughly analysed \cite{SpergelWMAP3}.) The value of $m_H^2$,
depends on the inflaton potential and is small with respect to $1$, 
but non-zero, in slow-roll inflation. 
As can be seen in FIG.~\ref{fig1}, the proposed physical two-point function
of Eq.\ (\ref{eq:DeltaSquared}) remains positive and approaches $0$ 
rapidly for large $v$, so the position-space two-point function is finite.
The spectra of Eqs.\ (\ref{eq:DeltaSquared-h}) 
and (\ref{eq:DeltaSquared-omega}) increase with $v$, and in the
absence of an explicit cut-off, lead to an infinite position-space 
two-point function. 

% Put a graph here of the spectra as a function of v for m = \sqrt{0.1},
% which is the first case in the table below. I had to use spectra.pdf
% as the file in the same directory as this tex file. When I tried to
% use spectra.eps it did not recognize the ending. So I opened the
% eps file in Preview by double clicking on it. Then I saved the result
% as a pdf file by just saving it from Preview. Then the graph showed
% up in the pdf of the paper. This is in contrast to the the RevTex4
% User's Guide, which said to use the eps file.

A notable feature of the proposed physical two-point function
spectrum is that the amplitude of the perturbations is significantly 
smaller than the value used in the literature. 
Earlier conclusions in the literature 
(for example, \cite{GuthPi82, BardeenSteinhardtTurner83})
that the amplitude predicted by certain theories (such as grand 
unified theories) was too large by several orders of magnitude 
will have to be reconsidered in light of the present
correction to the spectrum.
In TABLE~\ref{tab:spectra}, the values of each of the three spectra at 
horizon exit ($v=1$) are shown for four small values of $m^2/H^2$,
ranging from $0.1$ to $0.0001$. 
% Table : see LaTeX book page 64 and RevTex4 Guide
\begin{table}
\caption{Comparison of spectra at $v=1$}
\label{tab:spectra}
%\begin{ruledtabular}
\begin{tabular}{ l | c | c | c | c |}
                         & $\Delta_{\phi}^2$ & $\Delta_{\phi}^2{}_{\ \rm Hankel}$ & 
                                   $\Delta_{\phi}^2{}_{\ \rm \omega}$\\  \hline
  $m_H^2=0.1$ & $1.192\times 10^{-3} H^2$ & 
                               $4.868\times 10^{-2} H^2$ & 
                                 $2.415\times 10^{-2} H^2$\\  \hline
  $m_H^2=0.01$ & $1.147\times 10^{-4} H^2$ & 
                                 $5.046\times 10^{-2} H^2$ & 
                                    $2.520\times 10^{-2} H^2$\\  \hline
  $m_H^2=0.001$ & $1.138\times 10^{-5} H^2$ & 
                                 $5.064\times 10^{-2} H^2$ & 
                                    $2.532\times 10^{-2} H^2$\\  \hline
  $m_H^2=0.0001$ & $1.137\times 10^{-6} H^2$ & 
                                 $5.066\times 10^{-2} H^2$ & 
                                    $2.533\times 10^{-2} H^2$\\  \hline
\end{tabular}
%\end{ruledtabular}
\end{table}
We see that the proposed physical spectrum has an amplitude 
at horizon exit that decreases with decreasing $m$ and,
for the values of $m$ shown, is from one- to four-powers 
of $10$ smaller than the unrenormalized spectra.

It is not difficult to prove from Eq.\ (\ref{eq:DeltaSquared})
that when $m_H^2 = 0$, the subtraction terms give
the exact result that $\Delta_{\phi}^2(k,t) = 0$ for {\em all} $v$. 
This is a remarkable 
result, in view of the fact that without the adiabatic subtractions the 
massless minimally-coupled field has a {\em formal} two-point function
in position space that is infinite in the deSitter (Bunch-Davies)
vacuum state because of an IR divergence (as well as the UV
divergence). The adiabatic subtractions are determined by the need 
to cancel the UV divergences. There is no freedom to change them
at the IR end of the spectrum. The cancellation of 
the IR divergence by the same subtractions that tame the
UV divergence appears to be fortuitous, but may have a
deeper significance, perhaps pointing to a duality between 
IR and UV behavior.

 It was first shown by 
 Ford and Parker \cite[see discussion after Eq.\ (3.26)]{FordParker77},
 for the mode functions that define the Bunch-Davies de Sitter vacuum 
 state, that the formal two-point function of the massless, minimally-coupled 
 scalar field in this vacuum state has an infrared (IR) divergence.
 The existence of a similar IR divergence for the de Sitter vacuum
 state that has the full set of deSitter symmetries, including global discrete 
 space and time reflections on the de Sitter hypersphere, was found by 
 Allen \cite{Allen85}. This IR divergence has been taken as a reason to 
 reject the deSitter vacuum as a possible vacuum state of the
 minimally-coupled scalar field when $m=0$. But as we have seen,
 this IR divergence is cancelled and the {\em physical} two-point function of 
 this massless field in the Bunch Davies vacuum turns out to be exactly 
 zero instead of infinity. To our knowledge, this is  new result.
 
 Although, in slow-roll inflation $m$ is {\em not} $0$, the tensor 
 perturbations (gravitons) of each polarization that are created by 
 the inflationary expansion of the universe are governed in 
 the Lifschitz gauge by an equation like that of a 
 minimally-coupled $0$-mass inflaton.
 A gauge invariant analysis would be necessary to determine if
 the above result for the massless scalar field may be relevant to the 
 spectrum of tensor perturbations resulting from inflation.
 
\section{Conclusions}
We have shown that established methods of quantum field theory in 
curved spacetime give a major correction to the predicted amplitude
of the perturbation spectrum that results from inflation. This correction 
will be crucial in comparing theories that predict the 
magnitude of the Hubble parameter $H$ during inflation to the
beautiful observations of the cosmic microwave background and
the large scale structure of the universe.

\end{document}